\documentstyle[prc,preprint,aps]{revtex}
\begin{document}
\draft
\title{A particle-number-conserving solution to the 
generalized pairing problem}
\author{Feng Pan\footnote{On leave from the Department of Physics,
Liaoning Normal University, Dalian 116029, P. R.~China},
J. P. Draayer, and W. E. Ormand}
\address{Department of Physics and Astronomy, 202 Nicholson Hall,\\
Louisiana State University, Baton Rouge, LA 70803-4001}
\maketitle
\begin{abstract}
An exact, number-conserving solution to the generalized, orbit-dependent
pairing problem is derived by introducing an infinite-dimensional 
algebra. A method for obtaining eigenvalues and eigenvectors of 
the corresponding Hamiltonian is also given. 
The relevance of the orbit-dependent pairing solution is demonstrated by 
comparing predictions of the model with shell-model calculations.
\end{abstract}

Pairing has long been considered an important interaction in physics. The
concept can be traced back to the seniority scheme introduced by Racah 
in atomic physics~\cite{r:ref1}. Its physical significance was first realized 
in the study of superconductivity~\cite{r:ref2}. 
Following the suggestions of Bohr, Mottelson, and Pines~\cite{r:ref3}, the 
first detailed application of pairing in nuclei was made by 
Belyaev~\cite{r:ref4}.  
The concept has since been applied to other phenomena:  high T$_c$ 
superconductivity~\cite{r:ref5,r:ref6}, applications using the 
Hubbard model~\cite{r:ref7}, and  
pairing phenomena in liquids~\cite{r:ref8} and metal clusters~\cite{r:ref9}.

BCS methods have yielded major successes in studies of superconductivity. 
When applied to nuclei, however, one must remember that the number of
spectroscopically active nucleons is typically too small ($n\sim 10$) to support
underlying assumptions of the theory, namely $\delta n/n$ is not negligible, and
as a consequence particle-number nonconservation effects can enter and give rise
to spurious states, nonorthogonal solutions, etc.
Although this challenge can be partially remedied by making use 
of particle-number
projection techniques~\cite{r:ref10}, the elegance and simplicity of the BCS
method are then compromised. Secondly, an essential feature of pairing
correlations  are even-odd differences, which are driven mainly by Pauli
blocking. It is difficult to treat these differences in the 
BCS formalism because
different quasi-particle bases must be introduced for  different blocked levels.
After an investigation into the accuracy of the  BCS approximation, Kerman and
Lawson suggested that an exact diagonalization of the pairing Hamiltonian is
necessary~\cite{r:ref10}. Based on these observations, a
particle-number-conserving method for treating the pairing problem in
well-deformed nuclei was put forward~\cite{r:ref11}. The method uses a
configuration-energy truncation scheme and takes the strength of the pairing
interaction to be the same for all orbitals. 
Unfortunately, because of the deformation, each orbital can only accommodate a
single pair of particles and this limits the applicability of the theory. 
Complementary mean-field techniques with approximate Lipkin-Nogami number
projection and non-trivial pairing interactions~\cite{r:ref12,r:ref13} have also
been found to be a useful means for handling pairing correlations in such
systems~\cite{r:ref14,r:ref15,r:ref16}. However, exact shell-model solutions,
when feasible, even if only approximate, are the best way to probe the true
nature of many-body correlation effects. 

The purpose of this contribution is to give an exact, particle-number-conserving
solution to the generalized, orbit-dependent pairing problem. The method can be
applied to any number of pairs in any model space, even to cases where exact
diagonalization is computationally prohibitive.   

The generalized pairing Hamiltonian for spherical
nuclei can be written as
\begin{equation}
\hat{H}=
\sum_{jm}
\epsilon_{j}a^{\dagger}_{jm}a_{jm}-
\vert G\vert
\sum_{jj^{\prime}}c_{jj^{\prime}}S^{+}(j)S^{-}(j^{\prime}),
\label{e:e1}
\end{equation}
where the $\epsilon_{j}$ are single-particle energies and
$S^{\pm}(j)$ and $S^{0}(j)$ are the pairing operators for a 
single-$j$ shell defined by
\begin{eqnarray}
S^{+}(j)&=&\sum_{m>0}(-)^{j-m}a^{\dagger}_{jm}a^{\dagger}_{j-m},
\nonumber\\
S^{-}(j)&=&\sum_{m>0}(-)^{j-m}a_{j-m}a_{jm},\nonumber\\
S^{0}(j)&=&{1\over{2}}\sum_{m>0}(a^{\dagger}_{jm} a_{jm}+
a^{\dagger}_{j-m}
a_{j-m}-1)~=~{1\over{2}}(\hat{N}_{j}-\Omega_{j}),
\label{e:e2}
\end{eqnarray}
where $\Omega_{j}\equiv j+1/2$ is the maximum number of pairs in
the $j$-th
shell, $\hat{N}_{j}$ is the $j$-th shell particle number operator, and the
$c_{jj^{\prime}}$ measure the orbit-orbit pairing strength.
Note that Hamiltonians with degenerate single-particle energies
($\epsilon_{j}=\epsilon$ for all $j$-orbitals) and separable pairing 
strengths ($c_{jj^{\prime}}=c^{*}_{j}c_{j^{\prime}}$) are special cases of
the general theory. 

In general, for $N$  pairs, Eq.~(\ref{e:e1}) can be diagonalized with 
bases states that are products of the single-$j$ shell pairing wave 
functions:
\begin{equation}
\vert N\rangle=\sum_{k_{i}}B_{k_{1}k_{2}\cdots
k_{p}}S^{+~k_{1}}_{j_{1}}S^{+~k_{2}}_{j_{2}}S^{+~k_{3}}_{j_{3}}\cdots
S^{+~k_{p}}_{j_{p}}\vert 0\rangle,
\label{e:e3}
\end{equation}
where the summation is restricted so that 
\begin{equation}
\sum_{i=1}^{p}k_{i}=N,
\label{e:e4}
\end{equation}
and the $B_{k_{1}k_{2}\cdots k_{p}}$ are expansion coefficients 
to be determined. Here, $\vert 0\rangle$ is the pairing vacuum state 
that satisfying the condition
\begin{equation}
S^{-}_{j}\vert 0\rangle~=~0~~~{\rm for~ all}~~j.
\label{e:e5}
\end{equation}
The dimensionality of the Hamiltonian matrix in this basis increases 
rapidly with increasing $N$ and the number of shells $p$. Due to the Pauli
Principle, it is less than or equal to the dimension of the 
irreducible representation (irrep) $[N\dot{0}]$ of
the unitary group $U(p)$,
\begin{equation}
\dim~\leq~{(p+N-1)!\over{N!(p-1)!}}.
\label{e:e6}
\end{equation}
The equal sign in Eq.~(\ref{e:e6}) holds when all the single-$j$ shell 
pairing wave
functions in the summation of Eq.~(\ref{e:e3}) are Pauli allowed. 
From Eq.~(\ref{e:e6}), it is clear that the problem quickly
becomes intractable because there are no analytical expressions or recursion
relations for determining the $B_{k_{1}k_{2}\cdots k_{p}}$ coefficients.

As in the quasi-spin case~\cite{r:ref17}, we consider a slightly 
simpler Hamiltonian
\begin{equation}
\hat{H}=\epsilon\sum_{jm}a^{+}_{jm}a_{jm}-\vert G\vert
S^{+}_{0}S^{-}_{0},
\label{e:e7}
\end{equation}
that is, one with degenerate single-particle energies and a separable pairing
interaction,
\begin{equation}
S^{+}_{0}=\sum_{j}c^{*}_{j}S^{+}(j),~S^{-}_{0}=\sum_{j}c_{j}S^{-}(j)
\label{e:e8}
\end{equation}
with the coefficients $c_{i}$ satisfying the condition
\begin{equation}
\sum_{i}\vert c_{i}\vert^{2}=1.
\label{e:e9}
\end{equation}
This defines generalized pairing as proposed by Talmi~\cite{r:ref18}. The
separability
assumption, though strong, is physically motivated as it links the
pair-pair interaction strength to the individual pair formation 
probability. In the notation of Eq.~(\ref{e:e1}), 
$c_{jj^{\prime}}=c^{*}_{j}c_{j^{\prime}}$ with
$\vert c_{j}\vert^{2}$ giving the percentage of single-$j$ shell pairing in the
Hamiltonian. In what follows, the $c_{j}$ are taken to be real. 
Because the total number of particles is a good quantum number in a
number-conserving theory, the single-particle term in Eq.~(\ref{e:e7}) is a
constant and can be dropped without loss.   
Hence, Eq.~(\ref{e:e7}) reduces to
\begin{equation}
\hat{H}=-\vert G\vert S^{+}_{0}S^{-}_{0}.
\label{e:e10}
\end{equation}
To diagonalize this Hamiltonian, consider an algebra generated by
\begin{eqnarray}
S^{0}_{m}&=&\sum_{j} c_{j}^{2m}S^{0}(j),\nonumber \\
S^{\pm}_{m}&=&\sum_{j}c_{j}^{2m+1}S^{\pm}(j).
\label{e:e11}
\end{eqnarray}
It is easy to show that these generators satisfy the following
commutation relations:
\begin{eqnarray}
{[S^+_m,S^-_n]}&=&2S^0_{m+n+1},\nonumber \\
{[S^0_m,S^\pm_n]}&=&\pm S^\pm_{m+n}.
\label{e:e12}
\end{eqnarray}
Therefore, the $\{ S^{\mu}_{m},~\mu=0,+,-;~m=0,\pm 1, \pm 2,\cdots\}$ 
form an infinite-dimensional algebra, one that differs only slightly from a 
general Lie-algebra of the affine type without central extension.

The unique lowest-weight state of this algebra is simply the product of
the single-$j$ shell pairing vacua with arbitrary seniority quantum 
numbers. Therefore, it suffices to consider the total seniority zero case. 
The lowest-weight state satisfies
\begin{equation}
S^{-}_{m}\vert 0\rangle=0;~m=0,~\pm 1,~\pm 2,~\cdots,
\label{e:e13}
\end{equation}
\noindent and
\begin{equation}
S^{0}_{m}\vert 0\rangle=-{1\over{2}}\sum_{j}\vert c_{j}\vert^{2m}\Omega_{j}
\vert 0\rangle=
\Lambda_{m}\vert 0\rangle.
\label{e:e14}
\end{equation}
Furthermore, it can be proven that the eigenvectors of $\hat{H}$
for any $N$ and non-zero energy eigenvalue  can be written as
\begin{equation}
\vert N\rangle={\cal N} S^{+}_{0}S^{+}_{x_{1}}S^{+}_{x_{2}}\cdots 
S^{+}_{x_{N-1}}
\vert 0\rangle,
\label{e:e15}
\end{equation}
where $\cal{N}$ is a normalization constant and
\begin{equation}
S^{+}_{x_{i}}=\sum_{j}{c_{j}\over{1-c^{2}_{j}x_{i}}}S^{+}(j).
\label{e:e16}
\end{equation}

To obtain the variables $\{ x_{i};~i=1,2,\cdots,N-1\}$, Eq.~(\ref{e:e15}) 
can be expanded in terms
of $x_{i}$ around $x_{i}=0$,
\begin{equation}
\vert N\rangle={\cal N} \sum_{n_{i}}x_{1}^{n_{1}}x_{2}^{n_{2}}\cdots
x^{n_{N-1}}_{N-1}S^{+}_{0}S^{+}_{n_{1}}S^{+}_{n_{2}}\cdots S^{+}_{n_{N-1}}
\vert 0\rangle,
\label{e:e17}
\end{equation}
where  the $S^{+}_{n_{i}}$ are the Fourier-Laurent
coefficients in the expansion of $S_{x_{i}}^{+}$, namely
\begin{equation}
S^{+}_{n_{i}}={1\over{2\pi i}}\oint_{0}dx_{i}~x_{i}^{n_{i}}S^{+}_{x_{i}}.
\label{e:e18}
\end{equation}
Using Eq.~(\ref{e:e17}) and the commutation relations of Eq.~(\ref{e:e12}), 
it is easily shown that the $x_{i}$,
with $i=1,2,\cdots,$ $N-1$, satisfy the relations
\begin{equation}
-{1\over{2}}\sum^{p}_{j=1}\Omega_{j}c^{2}_{j}\alpha{1\over{1-\alpha
y_{i}c^{2}_{j}}}
={1\over{y_{i}}}+
\sum_{k\neq i}{1\over{y_{i}-y_{k}}},~~i=1,~2,\cdots,~N-1,
\label{e:e19}
\end{equation}
\noindent with
\begin{equation}
\sum^{N-1}_{i=1}{1\over{y_{i}}}=1,
\label{e:e20}
\end{equation}
\noindent where
\begin{equation}
y_{i}=x_{i}/\alpha,~~\alpha=-{2\over{h+2\Lambda_{1}}},~~h\equiv 
E/(-\vert G\vert).
\label{e:e21}
\end{equation}
We note that although these relations were derived for $x_{i}\approx 0$,
they are nonetheless valid in the entire complex plane except 
at the singularities in 
Eqs.~(\ref{e:e19}) and (\ref{e:e20}).
Therefore, the coefficients $x_{i}$ (i=1,~2,$\cdots$, N-1) and 
eigenvalues of the pairing energy $E\neq 0$ are simultaneously determined by 
the system of equations Eqs.~(\ref{e:e19}) and (\ref{e:e20}).

Similarly, the eigenvectors for $E=0$ can be expressed as
\begin{equation}
\vert N,~0\rangle={\cal N} S^{+}_{x_{1}}S^{+}_{x_{2}}\cdots S^{+}_{x_{N}}\vert
0\rangle.
\label{e:e22}
\end{equation}
Using the same technique as in the $E\neq 0$ case, it can be shown that
in this case the $x_{i}$ with $i=1,2,\cdots, N$, are determined by the
following set of equations
\begin{eqnarray}
\sum_{j}\Omega_{j}{c^{2}_{j}\over{1-x_{1} c^{2}_{j}}}&=&0~~~
{\rm for }~~N=1,\label{e:e23}\\
{1\over{2}}\sum_{j}\Omega_{j}{c^{2}_{j}\over{x_{i} c^{2}_{j}-1}}&=&
\sum_{k\neq i}{1\over{x_{i}-x_{k}}},~~~ 
i=1,~2,~\cdots,~N,~~{\rm for }~~N\geq 2.
\label{e:e24}
\end{eqnarray}

At this point, we reiterate that our solution is for a separable, 
orbit-dependent pairing interaction as defined by Eq.~(\ref{e:e7}), 
and not for the 
generalized pairing Hamiltonian of Eq.~(\ref{e:e1}). If the separability 
assumption is valid,
the solutions should be good approximations to the more general theory. 
The power
of the method lies in the fact that it gives the exact, 
particle-number-conserving
solution of the pairing problem for any number of pairs in any model space, even
for cases where full matrix diagonalizations cannot be carried out. 
Next we show, by example, that the separability 
assumption is consistent with ``realistic''
shell-model results and hence that theory can be used with confidence to study
many-particle correlation phenomena in pair-rich systems.  

As a test of the method, comparisons with shell-model calculations for like
particles interacting through a ``realistic'' generalized  pairing interaction
were carried out.  The shell-model calculations were performed within the
configuration space defined by the nuclear 
{\it ds}-shell ($0d_{5/2}$, $0d_{3/2}$,
and $1s_{1/2}$ orbitals). The generalized pairing Hamiltonian of 
Eq.~(\ref{e:e1})
was defined using the $J=0$ two-body matrix elements of the universal {\it
ds}-shell Hamiltonian of Wildenthal~\cite{r:ref19}. Since the number of
$J=0$  states for any two-particle system is small 
(three for the {\it ds}-shell), it is reasonable to define the 
generalized pair in Eq.~(\ref{e:e8}) as the wave
function of the shell-model ground state.  For our {\it ds}-shell
example this means the coefficients are: $c_{d_{5/2}}=0.570243$,
$c_{d_{3/2}}=0.626089$, and $c_{s_{1/2}}=0.531823$. The strength of
the interaction was then adjusted to reproduce the two-particle, ground-state
energy, that is,  $\vert G \vert = 3.12778$. With the Hamiltonian so defined,
solutions were obtained for $N=1$, 2, and 3 pairs of like
particles~\cite{r:foot1}. Shown in Fig.~\ref{fig:fig1} is a 
comparison between the eigenenergies obtained from the shell model (SM) 
and the separable, generalized pairing method.
For reference, the quasi-spin (QSA) approximation gives -6.389, -10.684, and
-12.788, respectively, for the $N=1$, 2, and 3 pair cases~\cite{r:foot2}. A
further measure of how well the generalized  pairing solutions correspond to the
shell-model results is illustrated in Table~\ref{tab:tab2} where the percent
overlap between the shell-model and generalized pairing wave functions are 
given. In all cases, the wave functions obtained with the method 
presented here have better than a 94\% overlap 
(96\% for the ground state) with the shell-model
states, which indicates that the assumption of separability is sound.

An exact, particle-number-conserving solution to the generalized, 
orbit-dependent pairing problem has been derived. A key 
feature of the derivation is the use of an
infinite-dimensional algebra, which is new in nuclear physics. 
Because the theory is number-conserving, it goes well-beyond what 
has been done even for the case of equal pairing strengths which 
includes to quasi-spin as a special limit. The relevance of the 
theory was shown by comparing eigenvalues and eigenvectors with
``realistic'' shell-model results.  

This work was supported by the National Science Foundation through Grant No.
9603006 and Cooperative Agreement No. EPS-9550481 which includes matching from
the Louisiana Board of Regents Support Fund. Additional support for WEO from
DOE contract DE--FG02-96ER40985 is also acknowledged. 

\newpage
\bibliographystyle{try}

\begin{thebibliography}{11} 

\bibitem{r:ref1} G. Racah, Phys. Rev. {\bf 62}, 438 (1942). 

\bibitem{r:ref2} J. Bardeen, L. N. Cooper, and J. R. Schrieffer, Phys. Rev. 
{\bf 108}, 1175 (1957).

\bibitem{r:ref3} A. Bohr, B. R. Mottelson, and D. Pines, Phys. Rev. {\bf 110}, 
936 (1958).

\bibitem{r:ref4} S. T. Belyaev, Mat. Fys. Medd. {\bf 31}, 11 (1959).

\bibitem{r:ref5} M. Randeria, J. M. Duan, and L. Y. Shieh, Phys. Rev. Lett. 
{\bf 62}, 981 (1989).

\bibitem{r:ref6} S. Schmitt-Rink, C. M. Verma, and A. E. Ruckenstein, Phys. 
Rev. Lett. {\bf 63}, 445 (1989).

\bibitem{r:ref7} C. N. Yang, Phys. Rev. Lett. {\bf 63}, 2144 (1989).

\bibitem{r:ref8} D. W. Cooper, J. S. Batchelder, and M. A. Taubenblatt, J. 
Coll. Int. Sci. {\bf 144}, 201 (1991).

\bibitem{r:ref9} M. Barranco, S. Hernandez, and R. J. Lombard, Z. Phys. 
{\bf D22}, 659 (1992).

\bibitem{r:ref10} A. K. Kerman and R. D. Lawson, Phys. Rev. {\bf 124}, 162 
(1961).

\bibitem{r:ref11} J. Y. Zeng and T. S. Cheng, Nucl. Phys. {\bf A405}, 1 (1983);
{\bf A411}, 49 (1984); {\bf A414}, 253 (1984).

\bibitem{r:ref12} H. J. Lipkin, Ann. Phys. {\bf 31}, 528 (1960).

\bibitem{r:ref13} Y. Nogami, Phys. Rev. {\bf 134}, B313 (1964); 
Y. Nogami and I. J. Zucker, Nucl. Phys. {\bf 60}, 203 (1964).

\bibitem{r:ref14} N. D. Dang, P. Ring, and R. Rossignoli, Phys. Rev.  
{\bf C47}, 606 (1993).

\bibitem{r:ref15} W. Satula, R. Wyss, and P. Magierski, Nucl. Phys. 
{\bf A578}, 45 (1994).

\bibitem{r:ref16} P. G. Reinhard, W. Nazarewicz, and  J. A. Maruhn, Phys. Rev.
{\bf C53}, 2776 (1993).

\bibitem{r:ref17} A. K. Kerman, Ann. Phys. {\bf 12}, 300 (1961).

\bibitem{r:ref18} I. Talmi, Nucl. Phys. {\bf A172}, 1 (1971).

\bibitem{r:ref19} B. H. Wildenthal, Prog. Part. Nucl. Phys. {\bf 11}, ed. 
D.~H.~Wilkinson, (Pegamon, Oxford, 1984) p.5. (The mass scaling factor
$(18/A)^{0.3}$ was not included in the present analysis.)

\bibitem{r:foot1} 
If a diagonalization of the two-particle system is not feasible, an alternative
procedure that work almost as well is to  determine the coefficients by the
relative weight of the diagonal matrix elements. With this choice, the
coefficients would be $c_{d_{5/2}}=0.766482$, $c_{d_{3/2}}=0.512492$, and
$c_{s_{1/2}}=0.714841$ with $\vert G \vert = 2.28278.$

\bibitem{r:foot2} 
Hamiltonian~(\ref{e:e7}) reduces to the quasi-spin approximation when all the
parameters $c_{j}$ are taken to be the same. In this case, the fully paired
(seniority zero) solution is well-known: eigenenergy 
$E = -\vert G \vert N(\Omega-N+1)$ 
with eigenstate 
$[\frac{(\Omega-N)!}{N!\Omega!}]^{1/2}(S^{+})^{N}\vert 0 \rangle$, 
where $N$ is the number of pairs, $\Omega = \sum_{j}\Omega_{j}$ and 
$S^{+}= \sum_{j}S^{+}(j)$.

\end{thebibliography}

\newpage

\begin{table}
\caption{Percent overlap between the eigenstates obtained 
with the separable, generalized pairing method and the
shell model as a function of the number of pairs, $N$.}
\begin{tabular} {crrr}
eigenstate $\backslash$ N & 1 & 2 & 3 \\
\tableline
1  & 96.2 & 99.4 & 97.8 \\
2  & 94.7 & 97.6 & 97.0 \\
3  & 99.4 & 97.2 & 94.3 \\
4  &      & 98.8 & 95.8 \\
5  &      & 99.2 & 96.8 \\
6  &      &      & 97.0 \\
\end{tabular}
\label{tab:tab2}
\end{table}

\newpage
\begin{figure}
\caption{Comparison between the spectra obtained from the shell model (SM) and
the separable, generalized pairing (SGP) method as a function of the 
number of pairs, N. The dashed lines indicate the correspondance between
the levels in the two models.}
\label{fig:fig1}
\end{figure}

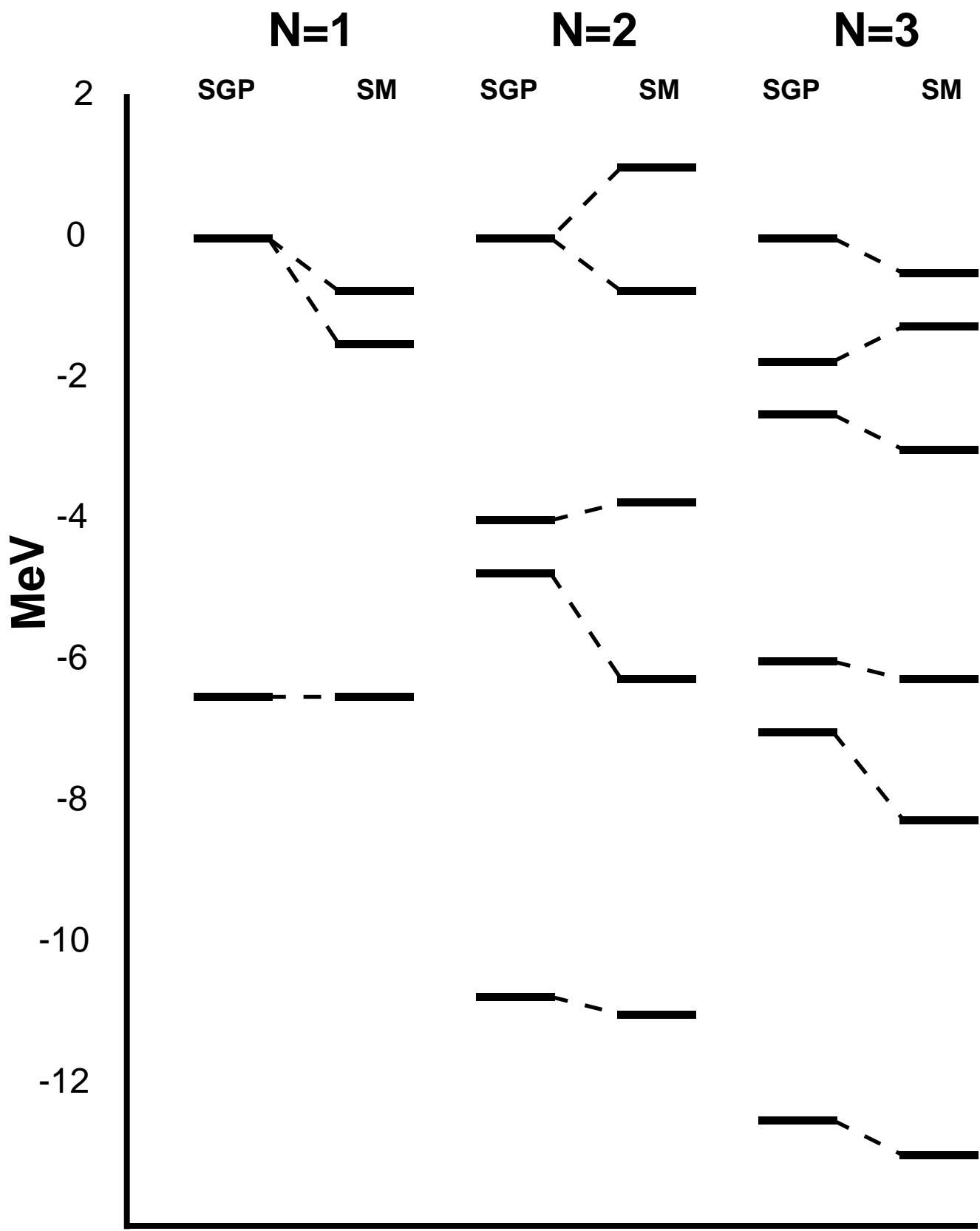
\end{document}